\documentclass[twoside]{article}
\usepackage{multicol}
\usepackage[a4paper, margin=1.5cm]{geometry}
\usepackage{graphicx,caption}% Include figure files
\usepackage{ragged2e} % Caption justification
\usepackage{dcolumn}% Align table columns on decimal point
\usepackage{bm}% bold math
\usepackage[utf8]{inputenc}
\usepackage[T1]{fontenc}
\usepackage{mathptmx}
\usepackage{amsmath}
\usepackage{etoolbox}
\usepackage[export]{adjustbox}
\usepackage{dblfloatfix}
\usepackage{physics}
\usepackage[table, svgnames, dvipsnames]{xcolor}
\usepackage{makecell, cellspace, caption}\usepackage[labelformat=empty]{caption}
\usepackage{hyperref}
\usepackage{soul}
\usepackage{lettrine}

\usepackage{xcolor} %define colors for table
\definecolor{mintbg}{rgb}{.63,.79,.95}
\colorlet{lightmintbg}{mintbg!30}
\colorlet{lightyellowgreen}{YellowGreen!20}
\usepackage{ctable} % for \specialrule command in table

\usepackage{array}  %Allows table and array manipulation
\newcolumntype{P}[1]{>{\centering\arraybackslash}p{#1}} %Generates a new type of column called P

\usepackage{authblk} %Author/affiliation package

\usepackage{fancyhdr}
\fancypagestyle{firststyle}
{
   \fancyhead[]{}
   \fancyfoot[RO]{\thepage}
}
\usepackage[maxbibnames = 5, url = false, doi = true, isbn=false,url=false,eprint=false]{biblatex} %Imports biblatex package
\hypersetup{
    colorlinks=true,
    linkcolor=blue,
    urlcolor=blue,
    citecolor=blue
}
\renewcommand{\thesection}{\Roman{section}.}

\usepackage{titlesec}

\titleformat{\section}  % which section command to format
  {\fontsize{11}{10}\bfseries\filcenter} % format for whole line
  {\thesection} % how to show number
  {0.5em} % space between number and text
  {} % formatting for just the text
  [] % formatting for after the text

\titlespacing\section{0pt}{10pt}{5pt}

\begin{document}

\title{\Large \textbf{Low loss hybrid Nb/Au superconducting resonators for \\ quantum circuit applications}}

\author[1]{\normalsize Marina Calero de Ory}
\author[1]{Victor Rollano\thanks{Corresponding author: vrollano@cab.inta-csic.es}}
\author[1]{David Rodriguez}
\author[1]{Maria Teresa Magaz}
\author[2]{Daniel Granados}
\author[1]{ \\ Alicia Gomez\thanks{Corresponding author: agomez@cab.inta-csic.es}} 

\affil[1]{Centro de Astrobiología (CSIC - INTA), Torrejón de Ardoz, 28850 Madrid, Spain.}
\affil[2]{IMDEA Nanociencia, Cantoblanco, Madrid 28049, Spain}
\date{}
\maketitle
\thispagestyle{firststyle}

\begin{center}
\parbox{17cm}{
Superconducting resonators are critical components in a wide range of quantum technologies, including high-performance detectors and amplifiers. However, their performance is often limited by noise and sensitivity issues caused by two-level systems (TLS) originated from oxide layers, such as Nb$_2$O$_5$, in the device. To address this challenge, we investigate a superconducting device combining a 100 nm niobium (Nb) circuit with a 10 nm gold (Au) capping layer. Our study explores the device’s performance across a wide range of temperatures and driving powers. Comparing with an uncapped device, we demonstrate that the Au capping layer effectively avoids oxide formation, thus reducing the TLS density and increasing the internal quality factors. Moreover, an increase in the non-linearity response is also observed. These findings provide valuable insights into improving resonator performance by addressing material-specific limitations, paving the way for more robust and efficient superconducting devices and highlight the potential of Nb/Au lumped element resonators for applications where TLS noise mitigation is critical.}
\end{center}

\medskip

\begin{multicols}{2}
\section{INTRODUCTION}Superconducting microwave resonators are key building blocks in the development of different technologies with a wide range of applications including quantum computing \autocite{blais2004}, quantum communication \autocite{rosario2023}, quantum sensing \autocite{baumer2019, baselmans2022} or cavities for particle accelerators \autocite{padamsee2014}. These superconducting resonators stand out for their high internal quality factor ($Q_i$) which essentially quantifies its electromagnetic (EM) energy loss rate; a higher $Q_i$ implies longer EM energy storage, enabling higher integration times and, thus, higher signal to noise ratio (SNR). Typically, the value of $Q_i$ is limited by different dissipative sources including radiation losses, superconducting vortices or residual surface resistance \autocite{gurevich2023, krasnok2023}. In the single photon and low-temperature limits, $Q_i$ is predominantly limited by the parasitic noise generated by two-level systems (TLSs) developed via trapped charges and electric dipoles naturally present in amorphous dielectric layers in contact with the superconductor, such as native oxides at its surface or the substrate/superconductor interface. TLSs couple to the resonator electric field \autocite{pappas2011,martinis2005}, inducing non-desired energy dissipation and decoherence, degrading, the device performance. Therefore, reducing or even eliminating TLSs is a must to attain better control of the microwave losses and boost the resonator performance.

In this sense, ongoing research continues into developing resonators with reduced TLS noise \autocite{bejanin2022}. Different strategies have been approached, including the optimization of the nanofabrication processes \autocite{barends2010reduced}, substrate choice \autocite{barends2009noise}, surface treatment \autocite{bruno2015}, resonator design \autocite{gao2008, mcrae2020a}, or the use of different superconducting materials \autocite{barends2010minimal, mcrae2020b}. Typical superconducting materials employed for developing high-quality resonators include niobium (Nb) or aluminum (Al), depending on the application. For instance, Al is widely used to develop Kinetic Inductance Detectors, state-of-the-art radiation detectors typically used for astrophysics experiments \autocite{baselmans2012}. On the other hand, Nb resonators present higher critical temperature ($T_c \sim 9$ K) and larger critical fields, leading to better stability against temperature and magnetic field fluctuations. This resilience makes them suitable for quantum applications requiring magnetic fields, opening the path for the implementation of new hybrid quantum architectures based on molecular spins qudits \autocite{rollano2022, chiesa2023} or magnonic excitations \autocite{tabuchi2014, pirro2021}. Even though these materials have proven to have very high performance, their internal quality factor is still limited by the noise generated by the TLSs present in the native oxide layer \autocite{verjauw2021}.

In this work, we propose using bilayer niobium/gold (Nb/Au) Lumped Element Resonators (LERs). The thin gold layer deposited on top of the niobium film prevents oxidation, enhancing surface stability and reducing the density of TLSs. Thus, the presence of the Au layer contributes to increase the internal quality factor and performance reliability of the resonators. Recent studies support this effect, showing that encapsulation improves the decay time in superconducting devices\autocite{alghadeer2022surface, bal2024systematic}. Beyond its role in inhibiting oxidation of the Nb surfaces, the Au layer also offers advantages in sensing and quantum technologies. On one hand, proximity effect produced by the Au capping layer tends to increase Lk due to a higher density of quasiparticles in the superconductor \autocite{gueron1996, hu2020}. This parameter relates to the inertia of the superconducting charge carriers when an alternating current is applied and, distinct from its geometric counterpart, it exhibits a strong dependence on current and temperature. This characteristic leads to enhanced responsivity in detectors \autocite{valenti2019}, allows for the tunability of resonance frequency \autocite{vissers2015} and is suitable for development of parametric amplifiers \autocite{chien2023}. The kinetic inductance contribution also increases the characteristic impedance and, therefore, the zero-point voltage fluctuations, making it an ideal platform for efficiently coupling microwave photons to small electric dipole moments \autocite{samkharadze2016}. On the other hand, the integration of molecular systems for the development of an hybrid quantum processor based on molecular-based qudits \autocite{chiesa2023} also remains a challenge. In fact, the presence of a gap between the resonator and the molecular ensemble crystal wastes the region of the resonator mode volume where the electromagnetic fields are more intense \autocite{rollano2022}. The Au capping layer can facilitate the integration of self-assembled monolayers (SAMs) via thiol groups, which serve as anchors for magnetic molecules\autocite{cornia2011, gabarro2023, coronado2005isolated, trasobares2023hybrid}. This configuration minimizes the gap, thereby maximizing the intensity of the microwave magnetic field near the resonator surface and enhancing the coupling strength with the molecules. Finally, the native oxide of the superconductor surface is a significant obstacle when applying these devices to applications involving scanning tunnelling microscopy (STM) \autocite{berti2023}. In this bilayer design, the Au layer improves the performance of the device in STM conditions, avoiding the native oxide, which, together with the possibility of SAMs deposition, opens the door to coherent control of spin qubits on surfaces using Nb superconducting resonators \autocite{wang2023universal, ghirri2011}. 

Here, the experimental demonstration of Nb/Au superconducting LERs and its comparison with bare Nb ones is presented. Microwave characterization shows a systematic increase in the internal quality factor, reaching up to $10^6$, and a reduction of the TLS density when adding the Au as a protective layer opening the possibility for its use in superconducting quantum circuits applications. The paper is organized as follows: The first section provides a detailed description of the devices and presents a statistical analysis of their internal quality factors to determine whether the Au layer contributes to the suppression of resonator losses. In the main section, we investigate the impact of thermal and driving power conditions on the physics of the devices, correlating the observed trends with the \unskip\parfillskip 0pt \par
\end{multicols}

\begin{figure}[h!]
\captionsetup{width=\textwidth, justification=Justified, labelformat=empty}
\centering
\includegraphics[width=0.95\textwidth]{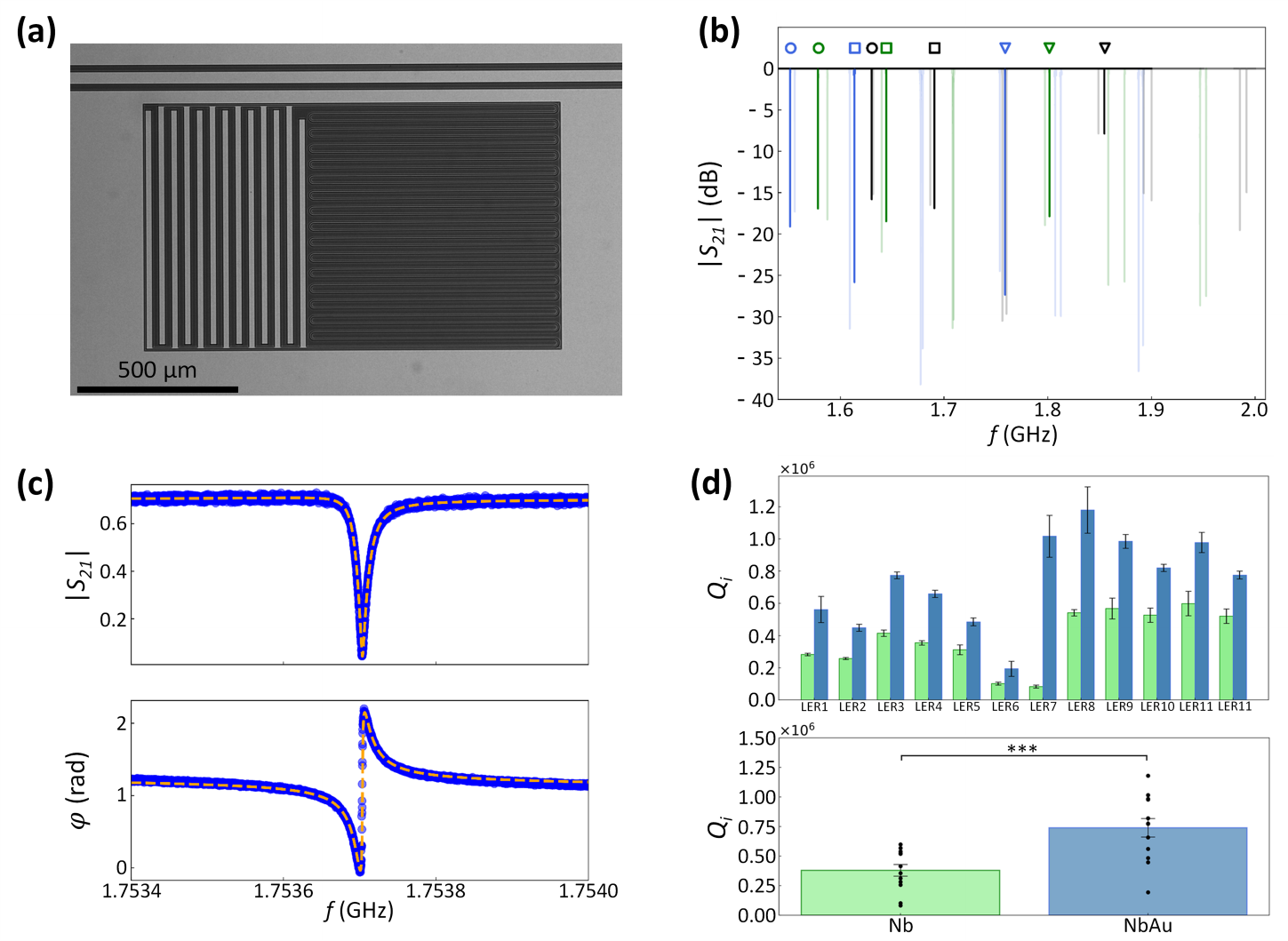}
\caption{\label{fig:fig1}FIG. 1. (a) Optical image of one of the fabricated Nb/Au LERs. (b) Amplitude of the transmission as a function of frequency measured for the twelve LERs on each chip. Measurements for the Nb and Nb/Au devices are shown in green and blue, respectively. The simulated resonances considering $L_k = 0$ pH sq\textsuperscript{-1} are depicted in black. Highlighted traces show the LERs resonances that are studied in detail for the rest of the article. Symbols above these resonances indicate which ones share the same LER geometry: LER1 design is marked by circles, LER4 by squares and LER8 by triangles.  These symbols will be consistently used in subsequent figures. (c) Amplitude (top) and phase (bottom) of the transmission as a function of frequency obtained for the Nb/Au LER8 resonator at 15 mK with a driving power of -96 dBm. Blue dots show the experimental data, while orange dashed lines show the fit obtained using Equation (\ref{eq:eq1}). Parameter values resulting from the fit are $f_r  = 1.7537$ GHz, $Q_i  = (1.18 \pm 0.14) \cdot 10^6$ and $Q_c= (7.90 \pm 0.08) \cdot 10^4$. (d) Top: Internal quality factor as a function of LER number for Nb (green) and Nb/Au (blue). Bottom: Results of a paired t-test performed on the two $Q_i$ distributions. The $Q_i$ values correspond to a driving power of -100 dBm. The analysis yields a p-value of 0.0002 (the three asterisks $***$ indicate that p $<$ 0.001).}
\end{figure}

\begin{multicols}{2}
\setlength{\parindent}{0em}characteristics of the TLSs present. Our study concludes with an examination of the non-linear response of the devices, followed by a final section that summarizes our findings and explores potential applications.

\bigskip
\section{RESULTS AND DISCUSSION}
\setlength{\parindent}{1.5em}A superconducting Nb/Au chip together with a nominally identical bare Nb chip as control sample have been developed. Each chip contains twelve LERs, which are coupled in parallel to a single coplanar waveguide transmission line (CPW) that serves for microwave readout. Resonators are labeled sequentially from LER1 to LER12, with higher labels corresponding to higher resonance frequency values. The LERs designs have been modeled through electromagnetic simulations using Sonnet Suite software \autocite{sonnet}. The resonance frequency is given by $f_r = 1/ 2 \pi \sqrt{LC}$, where $L$ is the inductance and $C$ the capacitance of each resonator. The CPW line is matched to a 50 $\Omega$ impedance, with a center conductor of 40 $\mu$m width separated 20 $\mu$m from the ground planes. The coupling quality factor ($Q_c$) is set by the distances between the LER and the CPW and ground planes. Figure \ref{fig:fig1}a shows an optical image of one fabricated Nb/Au LER as an example.

In Figure \ref{fig:fig1}b, we present the amplitude of the transmission parameter $|S_{21}|$ as a function of frequency for the twelve LERs on each chip. The data for Nb is shown in green and for Nb/Au in blue, both measured at 15 mK with a driving power of -96 dBm. Each resonance appears as a dip in the transmission spectrum. In this figure there are highlighted three resonances per chip, corresponding to resonators LER1, LER4, and LER8, indicating that their properties will be examined in detail later in the article. The three of them are designed with the same inductor, so their resonance frequency is set by changing the area of the capacitor, hence modifying the value of $C$.  For comparison, the simulated transmission considering a perfect conductor (with \( L_k = 0 \) pH sq\textsuperscript{-1}) is also plotted.

Figure \ref{fig:fig1}c illustrates the complex transmission spectrum (both amplitude, in the top, and phase, in the bottom) at 15 mK, driven by an input signal of -96 dBm for one of these three Nb/Au LERs as an example. The transmission parameter of a resonator can be modeled as follows:

\begin{equation} \label{eq:eq1}
S_{21} = a \left[ 1 - \frac{\frac{Q_l}{|Q_e|} e^{i\varphi}}{1 + 2i Q_l \left( \frac{f - f_r}{f_r}\right) } \right]
\end{equation}
where $Q_e = |Q_e|e^{i\varphi}$ is a complex-valued quality factor which is related to $Q_c$ through $Q_c^{-1} = Re(Q_e^{-1})$, with the imaginary part accounting for the resonance asymmetry. Then, the loaded quality factor ($Q_l$) is $Q_l^{-1}=Q_i^{-1}+Q_c^{-1}$. We fit the resonance data using the method detailed in \autocite{probst2015}, obtaining the resonator parameters $f_r$, $Q_i$, and $Q_e$. The fitting result is shown in Figure \ref{fig:fig1}c with orange dashed lines.

The comparison of the transmission spectra for Nb/Au and Nb LERs reveals a systematic shift to lower resonance frequencies ($f_r$) when adding the Au layer (see Figure \ref{fig:fig1}b). This behavior may be attributed to an excess of quasiparticles scattered at the boundary between the two materials, which leads to an increase in the kinetic inductance ($L_k$) \autocite{barends2009frequency}. From the experimental results, $L_k$ is determined for both materials by comparing the measured resonance frequencies ($f_{meas}$) with the simulated ones, using this expression:

\begin{equation} \label{eq:eq2}
L_{k} = \left[ \left(\frac{f_{sim}}{f_{meas}}\right)^2 - 1 \right] L_{g} 
\end{equation}
where $f_{sim}$ is the resonance frequency of the simulated resonator ($L_k = 0$ pH sq\textsuperscript{-1}) and $L_g$ is the geometric inductance, which is consistent across the three resonators due to their identical inductor lengths. The calculated mean value is $\Bar{L}_k = 0.13 \pm 0.01$ pH sq\textsuperscript{-1} for the Nb resonators and $\Bar{L}_k = 0.22 \pm 0.02$ pH sq\textsuperscript{-1} for the Nb/Au ones. Subsequently, the kinetic fraction ($\alpha_k$), defined as $L_k$ over the total inductance ($L_g + L_k$), can be obtained. Results show that Nb/Au resonators exhibit a higher $\alpha_k$, being the mean values $\Bar{\alpha}_k = 0.057 \pm 0.005$ for Nb and $\Bar{\alpha}_k = 0.094 \pm 0.005$ for Nb/Au. The parameter $\alpha_k$ is critical in developing kinetic inductance detectors; a higher $\alpha_k$ enhances device responsivity \autocite{day2003}. Additionally, a higher $\alpha_k$ leads to increased frequency tunability and amplification in superconducting devices developed for quantum applications \autocite{chien2023,vissers2015}.

A statistical analysis of the $Q_i$ values obtained from fitting the complex transmission measured at -100 dBm for the twelve LERs and both materials is presented\unskip\parfillskip 0pt \par
\end{multicols}

\begin{figure}[b]
    \captionsetup{width=\linewidth, justification=Justified, labelformat=empty}
    \includegraphics[width=1.0\linewidth]{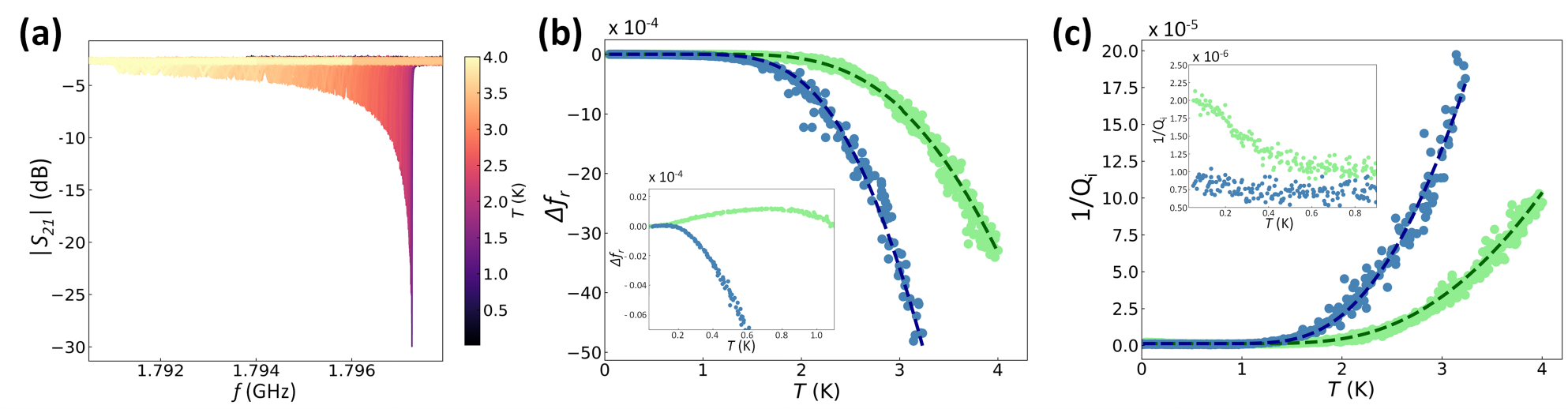}
    \captionof{figure}{FIG. 2. (a) Amplitude of the transmission spectrum in dB as a function of temperature for the Nb LER8 resonator measured with a driving power of -100 dBm. Color scale indicates the chip temperature. (b) Fractional change of the resonance frequency as a function of the chip temperature for a Nb/Au resonator (light blue) and a Nb resonator (light green). Dashed lines depict the Mattis-Bardeen fit upon the experimental data. Nb/Au fit is shown in dark blue while Nb fit is shown in dark green. Inset shows in detail the low-temperature regime. (c) Inverse of the internal quality factor as a function of temperature for Nb and Nb/Au. Color scheme is the same as in panel (b). Inset shows in detail the low-temperature regime.}
    \label{fig:fig2}
\end{figure}

\begin{multicols}{2}
\setlength{\parindent}{0em}in Figure \ref{fig:fig1}d. The upper plot compares each pair of LERs (corresponding to the same design), and illustrates that Nb/Au LERs consistently exhibit a higher internal quality factor, which may be due to a lower density of TLSs. We will further explore the origin of this difference in section 4. The lower plot contrasts the average $Q_i$ value for all LERs between the two materials. The paired t-test made upon the two groups of LERs provides a strong statistical evidence that there is a significant difference between the two groups of resonators when comparing the internal quality factors. From now on, we will analyze the physical properties of the aforementioned three LER designs (see Figure \ref{fig:fig1}b), focusing on the loss mechanisms that affect the internal quality factor of the resonators.

\setlength{\parindent}{1.5em}Temperature sweeps were performed for LER8, fabricated from both Nb and Nb/Au. Figure \ref{fig:fig2}a shows the amplitude of the transmission spectrum as a function of chip temperature ($T$), measured from 15 mK to 4 K, using the Nb LER8 as an example. The measurements were conducted at a low driving power (-100 dBm), where TLSs are not expected to saturate. As the temperature increases, the peak broadens and the resonance frequency shifts to lower values, reflecting the gradual degradation of the superconducting properties in the material. The peak broadening indicates increased resonator losses, while the resonance frequency shift results from the rise in kinetic inductance of the Cooper pairs within the superconducting material. By applying a fitting procedure using equation (\ref{eq:eq1}), we can determine $f_r$ and $Q_i$ as temperature increases.

Figure \ref{fig:fig2}b presents the fractional change in resonance frequency ($\Delta f_r$) for both materials as a function of temperature, calculated as $\Delta f_r (T) = (f_r(T) - f_{r,0}) / f_{r,0}$. Here, $f_{r,0}$ represents the resonance frequency at the lowest measured temperature (15 mK), where the superconducting properties of the device are fully restored. Figure \ref{fig:fig2}c shows the inverse of the internal quality factor as a function of temperature, which is proportional to the losses in the superconducting material. As observed in a study by Barends et al.\autocite{barends2009frequency}, the addition of an Au layer results in a superconductor whose parameters exhibit greater temperature dependence. This effect is attributed to the proximity effect, leading to a higher quasiparticle density at the interface between the two materials \autocite{burnett2017low}.

The ideal electromagnetic response of a superconductor is described by the Mattis-Bardeen (MB) model by using a complex conductivity $\sigma_s = \sigma_1 - i \sigma_2$ where the real part characterizes the dissipative processes while the imaginary part accounts for the inductive response of the superconductor. Specifically, $\sigma_1$ describes the behaviour of the quasiparticles while $\sigma_2$ relates to the Cooper pairs of the condensate. The model explains how $f_r$ and $Q_i$ depend on the complex conductivity:

\begin{equation} \label{eq:eq3}
    \Delta f_r ^{MB} (T) = \frac{\alpha_k}{2} \Delta \sigma_2 (T)
\end{equation}

\setlength{\parindent}{0em}and,

\begin{equation} \label{eq:eq4}
    Q_i^{-1} (T) = \alpha_k \frac{\sigma_1 (T)}{\sigma_2 (T)} 
\end{equation}

\setlength{\parindent}{1.5em}Figures \ref{fig:fig2}b and \ref{fig:fig2}c show in dashed lines the fits obtained from MB model using Equations (\ref{eq:eq3}) and (\ref{eq:eq4}). The analysis was performed assuming a temperature-dependent superconducting gap \autocite{mattis1958}. In the high-temperature regime, the onset of dissipative behaviour corresponds to $T \sim T_c / 5$, which is in good agreement with the MB model \autocite{zmuidzinas2012}. The values extracted from the fits for this LER are $\alpha_k (Nb) = 0.063 \pm 0.006$ with $T_c (Nb) = 8.7 \pm 0.2$ K, whereas $\alpha_k(Nb/Au) = 0.103 \pm 0.015$ with $T_c (Nb/Au) = 7.3 \pm 0.3$ K. These results are in good agreement with the $\alpha_k$ values obtained from the comparison made using equation (\ref{eq:eq2}). The reduction of $T_c$  and the increase of $\alpha_k$ observed in the Nb/Au can be attributed to the proximity effect.
 
As can be seen in the inset of Figure \ref{fig:fig2}b, in the low-temperature regime the behavior of the resonance frequency with temperature differs between the two materials. The interaction of the devices with the TLSs deviates the response of the superconductor from the MB behaviour. For the Nb device, $\Delta f_r$ exhibits an upshift as temperature rises, while for the Nb/Au resonator it presents a quick drop with no upshift. This trend in the resonance frequency translates into an increase in $Q_i^{-1}$ for Nb at low temperatures which is not observed in Nb/Au. The difference between the two materials may be consequence of the thermal saturation of the TLSs for $T_{ bath} = hf/2k_B$ \autocite{frasca2023nbn}. As temperature increases, the upper energy level of the TLSs gets populated, decreasing the losses of the resonator towards them and hence increasing $Q_i$. This is not observed for the Nb/Au resonator, for which $Q_i$ remains constant up to T $\sim 1.5$ K, revealing that microwave losses are not TLS limited in the case of Nb/Au.

\end{multicols}

\begin{figure}[b]
    \captionsetup{width=\textwidth, justification=Justified, labelformat=empty}
    \centering
    \includegraphics[width=0.99\linewidth]{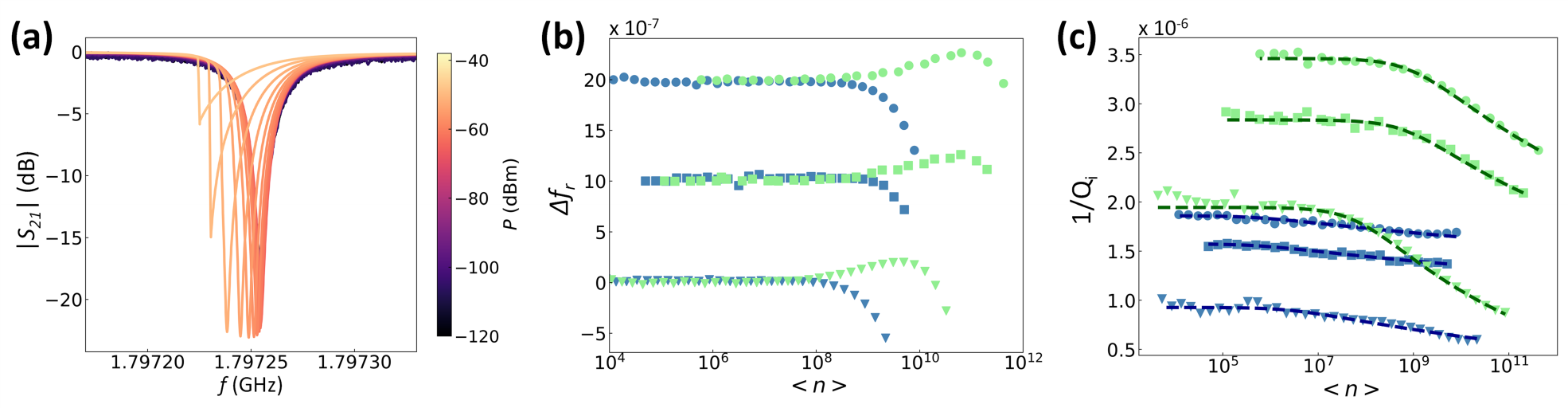}
    \caption{FIG. 3. (a) Transmission spectra as a function of driving power for the Nb LER8 resonator measured at 15 mK. (b) Fractional resonance frequency shift as a function of photon number for the Nb resonators (light green) and the Nb/Au ones (light blue). The photon number is obtained from Equation (\ref{eq:eq6}). Symbols represent the results for the different designs, circles for LER1, squares for LER4, and triangles for LER8. For clarity, the data corresponding to LER1 and LER4 has been upshifted +20 and +10 units respectivelly. (c) Inverse of the internal quality factor (tan ${\delta}$) as a function of the number of photons in the resonator. Experimental data in light green circles for Nb and light blue for Nb/Au. Dashed lines (dark green for Nb and dark blue for Nb/Au) show the fits performed using Equation (\ref{eq:eq7}) upon the experimental data.}
    \label{fig:fig3}
\end{figure}

%\end{multicols}

\begin{figure}[t]
    \captionsetup{width=\linewidth, justification=Justified,labelformat=empty}
    \includegraphics[width=0.95\linewidth]{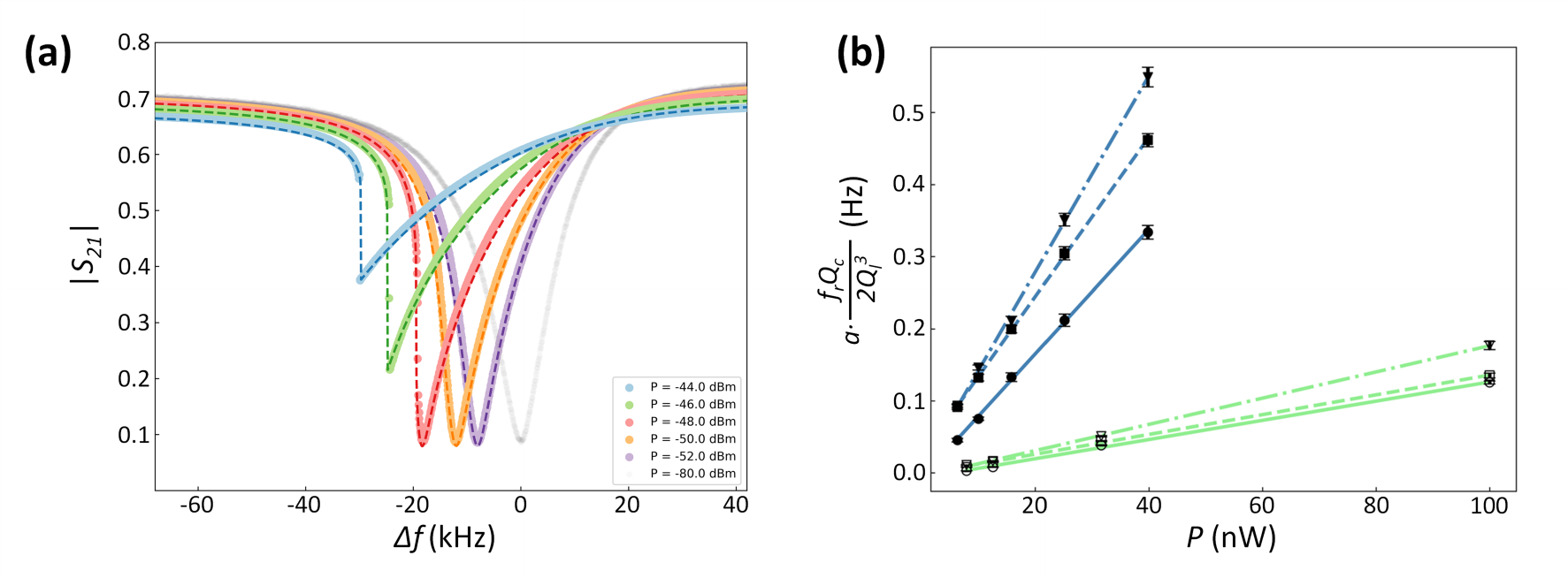}
    \captionof{figure}{FIG. 4. (a) Amplitude of the transmission as a function of frequency for the Nb/Au resonator LER1 in the non-linear regimen. Driving powers are between -44 dBm and -52 dBm. Dark-colored dashed lines correspond to the fit made using the non-linear model. Light gray curve represents the resonance data measured at low driving power (-80 dBm). The horizontal axis is referenced to the resonance frequency in the low-power regime ($f_{r,0}$). (b) Non-linearity parameter $a$ as a function of driving power (in linear scale) for the three LERs of both Nb and Nb/Au, along with the corresponding linear fit results in green and blue respectively. Symbols represent the results for the different designs, circles for LER1, squares for LER4 and triangles for LER8.}
    \label{fig:fig4}
\end{figure}

\begin{center}
\renewcommand{\arraystretch}{1.40}

\begin{tabular}{ l P{2.6cm} P{2.6cm} P{2.6cm} P{2.6cm} P{2.4cm}}
    \specialrule{.05em}{.0em}{.0em}
    \centering\textcolor{black}{\textbf{LER\#} (material)} &  
    \centering \(n_c\) & \(\beta \) & \(F \cdot \delta_{TLS}^0\) & \(Q_i^{sat}\) & \(f_r (GHz)\)\\
    \specialrule{.05em}{.0em}{.0em} %\rowcolor{lightyellowgreen}
    \textbf{LER1} (Nb) & \small $(1.02 \pm 0.11) \cdot 10^9$ & \small $(8.46 \pm 1.73) \cdot 10^{-2}$ & \small $(2.48 \pm 0.37) \cdot 10^{-6}$ & \small $(8.80 \pm 0.66) \cdot 10^5$ & \small 1.578422\\
    %\hline %\rowcolor{lightmintbg}
    \textbf{LER1} (Nb/Au)  & \small $(4.41 \pm 1.72) \cdot 10^5$ & \small $(3.05 \pm 1.31) \cdot 10^{-2}$ & \small $(8.77 \pm 4.05) \cdot 10^{-7}$ & \small $(9.64 \pm 1.21) \cdot 10^5$ & \small 1.551530\\
    \hline %\rowcolor{lightyellowgreen}
    \textbf{LER4} (Nb) & \small $(7.46 \pm 0.89) \cdot 10^8$ & \small $(1.14 \pm 0.25) \cdot 10^{-1}$ & \small $(1.81 \pm 0.34) \cdot 10^{-6}$ & \small $(8.72 \pm 0.67) \cdot 10^5$ & \small 1.644296\\
    %\hline %\rowcolor{lightmintbg}
    \textbf{LER4} (Nb/Au) & \small $(3.40 \pm 1.02) \cdot 10^6$ & \small $(4.81 \pm 2.46) \cdot 10^{-2}$ & \small $(6.07 \pm 2.46) \cdot 10^{-7}$ & \small $(9.99 \pm 1.15) \cdot 10^5$ & \small 1.613699\\
    \hline %\rowcolor{lightyellowgreen}
    \textbf{LER8} (Nb) & \small $(6.93 \pm 0.51) \cdot 10^7$ & \small $(1.15 \pm 0.12) \cdot 10^{-1}$ & \small $(1.97 \pm 0.59) \cdot 10^{-6}$ & \small $(9.52 \pm 0.69) \cdot 10^5$ & \small 1.797254\\
    %\hline %\rowcolor{lightmintbg}
    \textbf{LER8} (Nb/Au) & \small $(2.59 \pm 0.91) \cdot 10^6$ & \small $(4.65 \pm 2.58) \cdot 10^{-2}$ & \small $(9.74 \pm 4.08) \cdot 10^{-7}$ & \small $(9.91 \pm 1.09) \cdot 10^5$ & \small 1.753702\\
    \specialrule{.05em}{.0em}{.0em}
\end{tabular}

\captionof{table}{TABLE I. Summary of the parameters obtained from driving power characterization. Fit results from fitting the experimental data shown in Figure \ref{fig:fig4} using Equation (\ref{eq:eq7}). Last column shows the resonance frequency $f_r$ of each resonator obtained from fitting the complex trace using equation (\ref{eq:eq1}) in the low power regime with a driving power of -99 dBm. Error in resonance frequency is lower than $10^{-4}$ percent.}
\label{tab:tab1}
\end{center}

%\begin{multicols}{2}

\begin{multicols}{2}
To further investigate the nature of the loss mechanism in the devices, we have characterized the TLS contribution as a function of the driving power at 15 mK, where the thermal desaturation of the TLSs results in a power dependent resonator loss rate. As an example, Figure \ref{fig:fig3}a shows the transmission spectra of Nb LER 8 measured at different driving powers, ranging from -120 dBm to -40 dBm. From these values, the average number of photons in the resonator can be estimated using:

\begin{equation} \label{eq:eq6}
    \langle n \rangle = \frac{Q_l^2}{\pi Q_c} \frac{P_d}{h f_r^2}
\end{equation}

Figure \ref{fig:fig3}b shows $\Delta f_r$ as a function of the number of photons for the three LERs and both materials measured at 15 mK. For the Nb resonators we observe a positive resonance shift as $\langle n \rangle$ increases, which can be attributed to the gradual saturation of the TLSs. In the case of the Nb/Au resonators, this upshift is not observed which may be attributed to a low TLSs density. For higher values of $\langle n \rangle$, the nonlinearity of the kinetic inductance dominates the physics of the system when the current density ($J$) in the resonator reaches a value $J^*$, which sets the scale of the nonlinearity in the superconductor and it is close to the critical current density of the superconducting material \autocite{swenson2013}. As consequence, both LERs develop a strong asymmetric resonance response, which translates into a sharp jump in the transmission spectra and a negative relative frequency shift (see Figure \ref{fig:fig3}a as an example).

The inverse of the internal quality factor as a function of $\langle n \rangle$ at 15 mK is shown in Figure \ref{fig:fig3}c. Consistently, an overall enhancement of the $Q_i$ is obtained for the Nb/Au resonators in comparison with the Nb ones. Beyond a critical number of photons ($\langle n \rangle_c$), a steeper dependence of the microwave losses ($Q_i^{-1}$) with $\langle n \rangle$ is observed in the Nb LERs with respect the Nb/Au ones. This decrease in $Q_i^{-1}$ originates in the competition between the driving power and the relaxation ($\gamma_1$) and decoherence rates ($\gamma_2$) of the TLSs. The value of $\langle n \rangle_c$ is reached when the amplitude of the driving signal (i.e, the Rabi frequency $\Omega_R \sim \langle n \rangle^{1/2}$) increases beyond the TLSs loss rate $\sqrt{\gamma_1 \cdot \gamma_2}$, since more of them are being locked onto their excited state \autocite{andersson2021}. Hence, TLSs can absorb gradually less energy from the resonator and their contribution to the internal quality factor decreases. This behaviour can be modelized using the following expression \autocite{mcrae2020a}: 

\begin{equation} \label{eq:eq7}
    \frac{1}{Q_i} = \tan{\delta} = F \cdot \delta_{TLS}^0 \frac{\tanh{\left( \frac{h f_r}{2 k_B T}\right)}}{\left( 1 + \ \frac{\langle n \rangle}{\langle n \rangle_c}\right)^\beta} + \frac{1}{Q_i^{sat}}
\end{equation}
Here, $F$ is the ratio between the volume of the electric field coupled to the TLSs and the entire electric field volume created by the resonator, $\delta_{TLS}^0$ is the loss tangent associated to the TLSs density in the low temperature and photon limits, $\beta$ is an exponent that describes how strongly $Q_i$ depends on the photon number, and $1 / Q_i^{sat}$ is the high-power loss, for which the TLSs are saturated and other loss mechanisms dominate \autocite{mcrae2020b, burnett2017low}.

Dashed lines in Figure \ref{fig:fig3}c show the obtained fit made upon the experimental using the previous model. Table 1 compiles the values obtained from these fits. For the Nb resonators, the critical photon numbers are in the order of $10^8$, which are usually attributed to the contribution of the Nb$_2$O$_5$ oxide \autocite{verjauw2021}. Conversely, their Nb/Au counterparts show $\langle n \rangle_c$ values of about two orders of magnitude lower, which may be ascribed to the lower number of photons required to saturate the TLSs coupled to the capacitor due to the Au layer. Moreover, the loss tangent value ($ F \cdot \delta_{TLS}^0$) obtained for the Nb/Au resonators is systematically lower compared to the Nb ones. Finally, for both materials, a weak dependence of $Q_i^{-1}$ with the number of photons ($\beta \sim 0.1$) implies a strong TLS - TLS interaction, which is typical from Nb$_2$O$_5$ \autocite{verjauw2021, burnett2017low}; however, the analysis shows a weakened dependence induced by the presence of the Au layer, as evidenced by a smaller $\beta$ in Nb/Au resonators.

These results endorse the idea that Nb/Au resonators present a lower TLS density coupled to them, since the Au layer prevents the growth of Nb$_2$O$_5$. We supported this hypothesis by perfoming an energy dispersive X-ray (EDX) analysis in the capacitor fingers from both a capped and an uncapped resonator (LER8). Appendix B summarizes the results of the EDX analysis, proving that the oxygen concentration is higher in the uncapped device than in the capped one. The capacitor of the resonator is what couples electrically to the Nb$_2$O$_5$, so a lower oxide concentration implies lower TLSs density and hence a reduced critical number necessary to saturate the TLSs. Nevertheless, the critical photon numbers obtained for Nb/Au are not low enough to associate the resonators loss to the Si substrate native oxide. Under low driving power conditions, where TLS contributions are predominant, the Nb/Au resonators display internal quality factors that are approximately twice those observed in the Nb resonators. The $F \delta_{TLS}^0$ values for Nb, consistent with previous reports \autocite{gao2008}, exceed those for Nb/Au, indicating a reduced impact of TLSs in the latter. At the high-power regime, the values obtained for $Q_i^{sat}$ are similar for both materials, suggesting that the loss mechanism when the TLSs are saturated may have the same origin. 

Finally, we have investigated the effect of the Au layer on the kinetic inductance of the resonators by determining the\unskip\parfillskip 0pt \par
\end{multicols}

\begin{multicols}{2}
\setlength{\parindent}{1.5em}current density value in the inductor that marks the onset of non-linear behaviour ($J^*$). Figure \ref{fig:fig4}a shows several transmission spectra at high driving powers (between -44 dBm and -52 dBm) for the Nb/Au resonator LER1. As the current in the inductor increases, the resonance shifts towards lower frequencies developing a strong asymmetric profile with discontinuities (see also Figure \ref{fig:fig3}a). In the non-linear regime, the resonance model used to fit the experimental data is described in reference \autocite{swenson2013}. In this model, the parameter $a$ describes the strength of the non-linearity for a given driving power ($P_d$):

\begin{equation}
    a = \frac{2Q^3}{Q_c f_r} \frac{P_d}{E_*}
\end{equation}

\setlength{\parindent}{0em}where $E_*$ represents the energy scale at which the resonator physics starts to be dominated by the non-linearity.

\medskip
\begin{center}
\renewcommand{\arraystretch}{1.40}
\label{tab:tab2}
\begin{tabular}{ l P{2.4cm} P{2.4cm} P{2.4cm} P{2.4cm} }
    \specialrule{.05em}{.0em}{.0em}
    \centering\textcolor{black}{\small\textbf{LER\#} (material)} &  
    \centering \small$E_* (J)$ & \small$J_*(A \, cm^{-2})$ \\
    \specialrule{.05em}{.0em}{.0em} %\rowcolor{lightyellowgreen}
    \small\textbf{LER1} (Nb) & \footnotesize $(7.50 \pm 0.20) \cdot 10^{-7}$ & \footnotesize $(4.27 \pm 0.69) \cdot 10^{8}$ \\
    %\hline %\rowcolor{lightmintbg}
    \small\textbf{LER1} (Nb/Au)  & \footnotesize $(1.16 \pm 0.02) \cdot 10^{-7}$ & \footnotesize $(2.10 \pm 0.16) \cdot 10^{8}$ \\
    \hline %\rowcolor{lightyellowgreen}
    \small\textbf{LER4} (Nb) & \footnotesize $(7.28 \pm 0.14) \cdot 10^{-7}$ & \footnotesize $(4.21 \pm 0.68) \cdot 10^{8}$ \\
    %\hline %\rowcolor{lightmintbg}
    \small\textbf{LER4} (Nb/Au) & \footnotesize $(9.03 \pm 0.08) \cdot 10^{-8}$ & \footnotesize $(1.85 \pm 0.14) \cdot 10^{8}$ \\
    \hline %\rowcolor{lightyellowgreen}
    \small\textbf{LER8} (Nb) & \footnotesize $(5.49 \pm 0.07) \cdot 10^{-7}$ & \footnotesize $(3.65 \pm 0.59) \cdot 10^{8}$ \\
    %\hline %\rowcolor{lightmintbg}
    \small\textbf{LER8} (Nb/Au) & \footnotesize $(7.33 \pm 0.13) \cdot 10^{-8}$ & \footnotesize $(1.67 \pm 0.13) \cdot 10^{8}$ \\
    \specialrule{.05em}{.0em}{.0em}
\end{tabular}
\end{center}
\captionof*{table}{TABLE II. Fit results obtained from the non-linear regime analysis upon the experimental data shown in Figure 4a.}

\medskip

\setlength{\parindent}{1.5em}From fitting the transmission spectra in the non-linear regime we obtained the parameter $a$ as a function on $P_d$. These non-linear fits were performed for the three LER designs made from Nb/Au and Nb. The result is shown in Figure \ref{fig:fig4}b, where each data point comes from a non-linear fit. A systematic difference between Nb and Nb/Au has been found. The slope of this linear relation gives the value of $E_*^{-1}$. The results of our analysis are summarized in Table 2, showing that the onset of the non-linear regime occurs at lower energies for the Nb/Au LERs than for the Nb ones. The parameters $E_*$ and $J_*$ are related through $E_* \propto L_k J_*^2 / \alpha_k$. The values of $L_k$ and $\alpha_k$ we have used in this calculation are the ones given by the analysis made using Equation (\ref{eq:eq2}). The reduction found in $J_*$ originates from the excess of quasiparticles introduced by the Au layer, adding additional inertia into the system and hence lowering $J_*$. Since $L_k (J) \propto J / J_*$, the reduction in $J_*$ compared to the bare Nb resonators leads to an increase in the kinetic inductance value at a given density current.

\section{CONCLUSIONS}\setlength{\parindent}{1.5em}In summary, we have fabricated superconducting lumped element resonators made of an Nb/Au bilayer and characterized their properties (resonance frequency and internal quality factor) as a function of temperature and driving power. We have compared our results by performing the same experiments in a nominally identical control sample fabricated of bare Nb. The microwave characterization in our Nb/Au and Nb devices matches our simulated designs in a consistent manner, proving a reliable manufacturing process. Adding the Au layer avoids Nb oxidation, reducing TLS loss and thus increasing the internal quality factor of the resonator. In our study, we obtain internal quality factors higher than $10^6$ for the Nb/Au devices even when the TLSs are not saturated, proving that at this temperature range (15 mK), the Au capping layer does not spoil the resonator properties but improves them. Furthermore, the presence of the Nb/Au interface results in an excess of quasiparticles, increasing the kinetic inductance of the device. Overall, our results show that Nb/Au superconducting LERs outperform the bare Nb ones with reference to the internal quality factor, TLS contribution and responsivity with a direct impact in the applicability of these resonators. In particular, these devices may allow an optimized integration of molecular spin-based quantum devices thanks to the ease of functionalizing the Au layer and open new possibilities in STM experiments. In addition, the higher kinetic inductance and an increase nonlinearity in the Nb/Au devices make them more suitable to be applied as kinetic inductance parametric amplifiers, tunable resonators or single photon detectors.

\bigskip
\section*{APPENDIX A: FABRICATION DETAILS}The device fabrication process starts with the surface preparation of a 270 $\mu$m high-resistivity silicon (Si) substrate ($R > 1$  k$\Omega \cdot $ cm), using a 1\% hydrofluoric acid bath to remove the native silicon oxide from the wafer. In less than one minute after the acid bath, the sample is introduced in an electron-beam evaporator where the three layers, titanium (Ti) 2.5 nm/ Nb 100 nm / Au 10 nm, are deposited without a vacuum break in between. The thickness of the Au layer is set to 10 nm to ensure a continuous film. Similar procedure is used for the Ti 2.5 nm/ Nb 100 nm. In both cases, the Ti seed layer is used to improve the adhesion to Si substrate. The evaporation was performed after reaching a chamber pressure of 1$\cdot$10$^{-9}$ mbar while the evaporation pressure is maintained around 1$\cdot$10$^{-5}$ mbar to reach a target evaporation rate of 2 ${\textrm{\r{A}}}$/s for Ti and Au and 1 ${\textrm{\r{A}}}$/s for Nb. The patterning process is based on maskless laser-writer photolithography using AZnLOF 2070 negative photoresist. Afterwards, the etching process is performed in three steps: first, the gold film is etched by argon ion milling (rate 3.3 ${\textrm{\r{A}}}$/s), then the niobium film is removed using a SF6:Ar plasma etching (rate 14.3 ${\textrm{\r{A}}}$/s) and, finally, the Ti is removed using wet etching. The resist covering the device is removed with an RF oxygen plasma for 5 minutes, peeling the superficial layer burned during RIE. During plasma, the RF power is 50 W with an oxygen flow of 150 mL/min. The remaining resist is finally removed in an acetone and isopropanol rinse.

The devices are cooled down to milikelvin temperatures using a dilution refrigerator (Bluefors LD250), while the electrodynamic response is measured with a commercial Vector Network Analyzer (VNA). The input power is varied from – 60 dBm to + 20 dBm from the VNA and attenuated by 30 dB at room temperature, 20 dB in the 4 K stage and 10 dB in the 100 mK stage. DC blocks prevent DC current leakages and provide thermal breaks between stages. The output signal is then amplified by a low-noise amplifier at the 4 K stage and a second amplification stage outside the refrigerator. Stainless steel (SS), CuNi and (superconducting) NbTi cables are used between stages in order to guarantee good electrical conductivity and poor thermal conductivity, whilst copper cables are used to thermalize within the stage. The final devices are mounted in a cold finger in the mixing chamber plate on a superconducting aluminum holder with a PCB, using low temperature varnish (SCBltv01) for thermalization and aluminum wire-bonding for electrical connections. The sample is further magnetically screened using a mu-metal shield at room temperature.

\bigskip
\section*{APPENDIX B: EDX ANALYSIS}
The oxygen concentration in the capped and uncapped devices has been assessed using energy dispersive X-ray (EDX) in a Zeiss Sigma Scanning Electron Microscope (SEM) equipped with a Bruker X Flash 430 detector. The EDX mappings and spectra were obtained with a working distance of 8.4 mm and an accelerating voltage of 8.0 keV. The spectra are normalized to the intensity of the peak appearing at zero keV. figures B1a and B1b present the EDX mappings of a capacitor finger from LER8, utilizing the oxygen K$\alpha$ line at 0.525 keV to visualize the oxygen distribution within the sample. These mappings provide a spatial representation of oxygen concentration, highlighting with a white rectangle the capacitor regions at which the EDX spectra were recorded. While in the spectrum obtained from the uncapped capacitor, the O K$\alpha$ peak stands out from the bremsstrahlung background, this peak is absent in the spectrum obtained from the gold-capped capacitor, indicating the lower oxygen concentration. The inset in figure B1c focuses on the O K$\alpha$ line, allowing for a detailed examination of the oxygen signal. 

\begin{center}
    \captionsetup{width=\linewidth, justification=Justified,labelformat=empty}
    \includegraphics[width=0.975\linewidth]{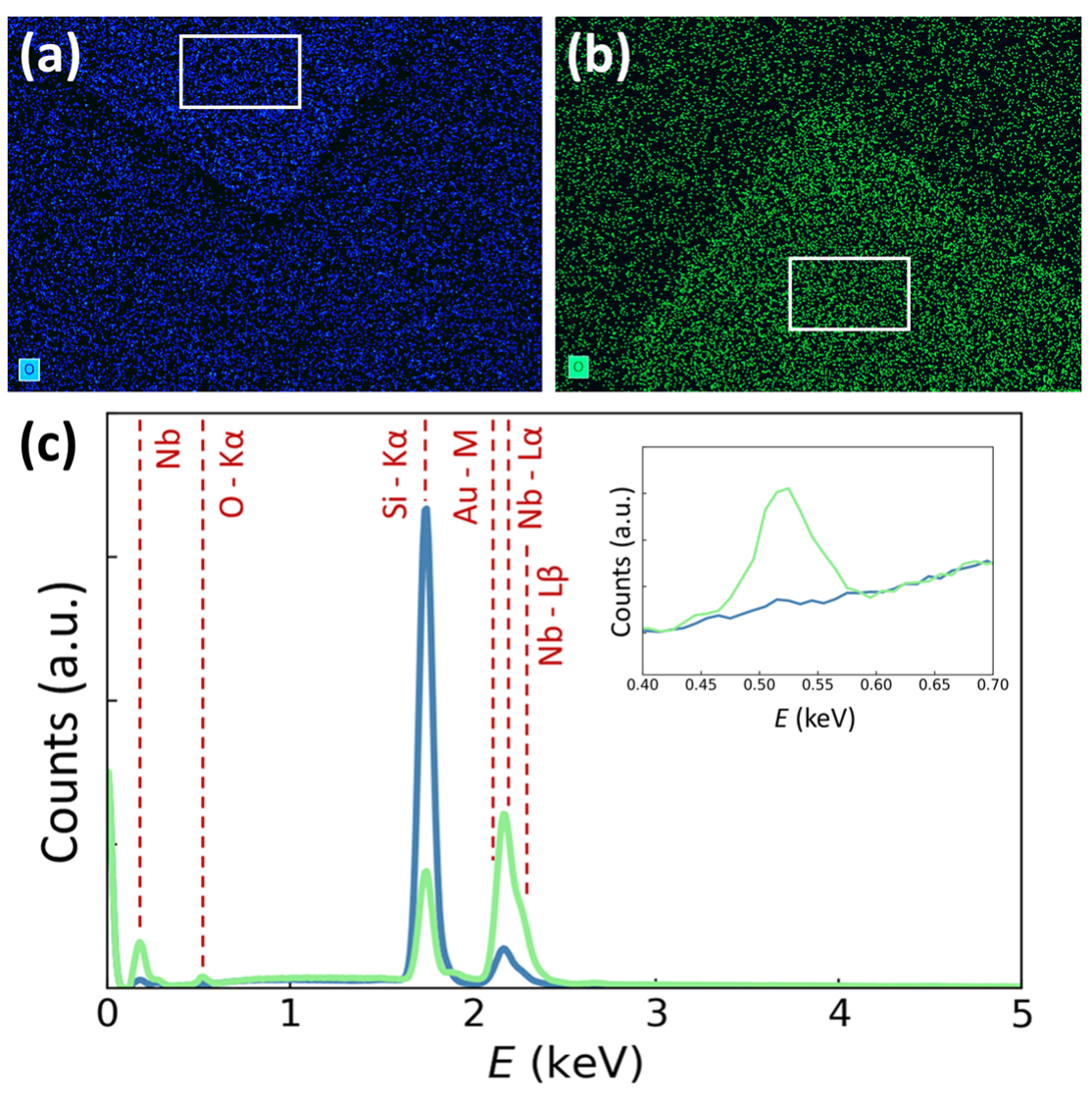}
    \captionof{figure}{FIG B1. (a) EDX mapping using the Oxygen K$\alpha$ line at 0.525 keV for an uncapped device. The white square over the resonator capacitor indicates the area from which the EDX spectrum has been acquired. (b) EDX mapping using the Oxygen K$\alpha$ line at 0.525 keV for a capped device. The white square over the resonator capacitor indicates the area from which the EDX spectrum has been acquired. (c) Comparison of the two EDX spectra from Nb/Au (blue) and Nb (green) samples. These spectra were acquired from areas indicated by the white squares in the EDX mappings. The different peaks in the spectra have been indicated using red dahsed lines. The inset shows a narrow energy band centered on the O K$\alpha$ line.}
    
    \label{fig:figb1}
\end{center}

\section*{ACKNOWLEDGEMENTS} \par %delete if not applicable))
This work has received support from grants TED2021-131447B-C22, PID2022-137779OB-C41, PID2022-137779OB-C42 and the research network RED2022-134839-T funded by the Spanish MCIN/AEI/10.13039/501100011033, by the EU “NextGenerationEU”/PRTR and by the “ERDF A way of making Europe”. IMDEA Nanoscience acknowledges financial support from the “Severo Ochoa” Programme for Centres of Excellence in R\&D (CEX2020-001039-S) and CAB from the the CSIC Research Platform PTI-001 and from “Tecnologías avanzadas para la exploración del Universo y sus componentes” (PR47/21 TAU-CM) project funded by Comunidad de Madrid, by “NextGenerationEU”/PRTR.

\AtNextBibliography{\footnotesize}
\printbibliography[heading=none]

@article{blais2004,
  title = {Cavity quantum electrodynamics for superconducting electrical circuits: An architecture for quantum computation},
  author = {Blais, Alexandre and Huang, Ren-Shou and Wallraff, Andreas and Girvin, S. M. and Schoelkopf, R. J.},
  journal = {Phys. Rev. A},
  volume = {69},
  pages = {062320},
  numpages = {14},
  year = {2004},
  publisher = {American Physical Society},
  doi = {10.1103/PhysRevA.69.062320},
  url = {https://link.aps.org/doi/10.1103/PhysRevA.69.062320}
}

@article{rosario2023,
  title = {Collateral Coupling between Superconducting Resonators: Fast High-Fidelity Generation of Qudit-Qudit Entanglement},
  author = {Rosario, Pedro and Santos, Alan C. and Villas-Boas, C.J. and Bachelard, R.},
  journal = {Phys. Rev. Appl.},
  volume = {20},
  pages = {034036},
  numpages = {11},
  year = {2023},
  publisher = {American Physical Society},
  doi = {10.1103/PhysRevApplied.20.034036},
  url = {https://link.aps.org/doi/10.1103/PhysRevApplied.20.034036}
}

@article{baumer2019,
  title={Quantitative modeling of superconducting planar resonators for electron spin resonance},
  author={Weichselbaumer, Stefan and Natzkin, Petio and Zollitsch, Christoph W and Weiler, Mathias and Gross, Rudolf and Huebl, Hans},
  journal={Physical Review Applied},
  volume={12},
  number={2},
  pages={024021},
  year={2019},
  publisher={APS},
  doi = {10.1103/PhysRevApplied.12.024021}
}

@article{baselmans2022,
  title={Ultra-sensitive THz microwave kinetic inductance detectors for future space telescopes},
  author={Baselmans, J.J.A. and Facchin, F. and Pascual-Laguna, A. and Bueno, J. and Thoen, D.J. and Murugesan, V. and Llombart, N. and De Visser, P.J.},
  journal={Astronomy \& Astrophysics},
  volume={665},
  pages={A17},
  year={2022},
  publisher={EDP Sciences},
  doi = {10.1051/0004-6361/202243840}
}

@article{rollano2022,
  title={High cooperativity coupling to nuclear spins on a circuit quantum electrodynamics architecture},
  author={Rollano, Victor and de Ory, Marina C and Buch, Christian D and Rub{\'\i}n-Osanz, Marcos and Zueco, David and S{\'a}nchez-Azqueta, Carlos and Chiesa, Alessandro and Granados, Daniel and Carretta, Stefano and Gomez, Alicia and others},
  journal={Communications Physics},
  volume={5},
  number={1},
  pages={246},
  year={2022},
  publisher={Nature Publishing Group UK London},
  doi = {10.1038/s42005-022-01017-8}
}

@article{padamsee2014,
  title={Superconducting radio-frequency cavities},
  author={Padamsee, Hasan S},
  journal={Annual review of nuclear and particle science},
  volume={64},
  pages={175--196},
  year={2014},
  publisher={Annual Reviews},
  doi = {10.1146/annurev-nucl-102313-025612}
}

@article{gurevich2023,
  title={Tuning microwave losses in superconducting resonators},
  author={Gurevich, Alexander V},
  journal={Superconductor Science and Technology},
  volume={36},
  number={6},
  pages={063002},
  year={2023},
  publisher={IOP Publishing},
  doi = {10.1088/1361-6668/acc214}
}

@article{krasnok2023,
  title={Advancements in Superconducting Microwave Cavities and Qubits for Quantum Information Systems},
  author={Krasnok, Alex and Dhakal, Pashupati and Fedorov, Arkady and Frigola, Pedro and Kelly, Michael and Kutsaev, Sergey},
  journal={arXiv preprint arXiv:2304.09345},
  year={2023},
  doi = {10.48550/arXiv.2304.09345}
}

@article{pappas2011,
  title={Two level system loss in superconducting microwave resonators},
  author={Pappas, David P and Vissers, Michael R and Wisbey, David S and Kline, Jeffrey S and Gao, Jiansong},
  journal={IEEE Transactions on Applied Superconductivity},
  volume={21},
  number={3},
  pages={871--874},
  year={2011},
  publisher={IEEE},
  doi = {10.1109/TASC.2010.2097578}
}

@article{martinis2005,
  title={Decoherence in Josephson qubits from dielectric loss},
  author={Martinis, John M and Cooper, Ken B and McDermott, Robert and Steffen, Matthias and Ansmann, Markus and Osborn, KD and Cicak, Katarina and Oh, Seongshik and Pappas, David P and Simmonds, Raymond W and others},
  journal={Physical review letters},
  volume={95},
  number={21},
  pages={210503},
  year={2005},
  publisher={APS},
  doi = {10.1103/PhysRevLett.95.210503}
}

@article{bejanin2022,
  title={Fluctuation spectroscopy of two-level systems in superconducting resonators},
  author={B{\'e}janin, JH and Ayadi, Y and Xu, X and Zhu, C and Mohebbi, HR and Mariantoni, M},
  journal={Physical Review Applied},
  volume={18},
  number={3},
  pages={034009},
  year={2022},
  publisher={APS},
  doi = {10.1103/PhysRevApplied.18.034009}
}

@article{barends2010reduced,
    author = {Barends, R. and Vercruyssen, N. and Endo, A. and de Visser, P. J. and Zijlstra, T. and Klapwijk, T. M. and Baselmans, J. J. A.},
    title = "{Reduced frequency noise in superconducting resonators}",
    journal = {Applied Physics Letters},
    volume = {97},
    number = {3},
    pages = {033507},
    year = {2010},
    month = {07},
    issn = {0003-6951},
    doi = {10.1063/1.3467052},
}

@ARTICLE{barends2009noise,
  author={Barends, R. and Hortensius, H. L. and Zijlstra, T. and Baselmans, J. J. A. and Yates, S. J. C. and Gao, J. R. and Klapwijk, T. M.},
  journal={IEEE Transactions on Applied Superconductivity}, 
  title={Noise in NbTiN, Al, and Ta Superconducting Resonators on Silicon and Sapphire Substrates}, 
  year={2009},
  volume={19},
  number={3},
  pages={936-939},
  doi={10.1109/TASC.2009.2018086}}

@article{bruno2015,
    author = {Bruno, A. and de Lange, G. and Asaad, S. and van der Enden, K. L. and Langford, N. K. and DiCarlo, L.},
    title = "{Reducing intrinsic loss in superconducting resonators by surface treatment and deep etching of silicon substrates}",
    journal = {Applied Physics Letters},
    volume = {106},
    number = {18},
    pages = {182601},
    year = {2015},
    month = {05},
    issn = {0003-6951},
    doi = {10.1063/1.4919761},
    url = {https://doi.org/10.1063/1.4919761},
    eprint = {https://pubs.aip.org/aip/apl/article-pdf/doi/10.1063/1.4919761/13955116/182601\_1\_online.pdf},
}

@article{gao2008,
    author = {Gao, Jiansong and Daal, Miguel and Vayonakis, Anastasios and Kumar, Shwetank and Zmuidzinas, Jonas and Sadoulet, Bernard and Mazin, Benjamin A. and Day, Peter K. and Leduc, Henry G.},
    title = "{Experimental evidence for a surface distribution of two-level systems in superconducting lithographed microwave resonators}",
    journal = {Applied Physics Letters},
    volume = {92},
    number = {15},
    pages = {152505},
    year = {2008},
    month = {04},
    issn = {0003-6951},
    doi = {10.1063/1.2906373},
    url = {https://doi.org/10.1063/1.2906373},
    eprint = {https://pubs.aip.org/aip/apl/article-pdf/doi/10.1063/1.2906373/14393210/152505\_1\_online.pdf},
}

@article{mcrae2020a,
  title={Dielectric loss extraction for superconducting microwave resonators},
  author={McRae, Corey Rae H and Lake, RE and Long, JL and Bal, Mustafa and Wu, Xian and Jugdersuren, Battogtokh and Metcalf, TH and Liu, Xiao and Pappas, David P},
  journal={Applied Physics Letters},
  volume={116},
  number={19},
  year={2020},
  publisher={AIP Publishing},
  doi = {10.1063/5.0004622}
}

@article{barends2010minimal,
    author = {Barends, R. and Vercruyssen, N. and Endo, A. and de Visser, P. J. and Zijlstra, T. and Klapwijk, T. M. and Diener, P. and Yates, S. J. C. and Baselmans, J. J. A.},
    title = "{Minimal resonator loss for circuit quantum electrodynamics}",
    journal = {Applied Physics Letters},
    volume = {97},
    number = {2},
    pages = {023508},
    year = {2010},
    month = {07},
    issn = {0003-6951},
    doi = {10.1063/1.3458705},
    url = {https://doi.org/10.1063/1.3458705},
    eprint = {https://pubs.aip.org/aip/apl/article-pdf/doi/10.1063/1.3458705/14434537/023508\_1\_online.pdf},
}

@article{mcrae2020b,
  title={Materials loss measurements using superconducting microwave resonators},
  author={McRae, Corey Rae Harrington and Wang, Haozhi and Gao, Jiansong and Vissers, Michael R and Brecht, Teresa and Dunsworth, Andrew and Pappas, David P and Mutus, Josh},
  journal={Review of Scientific Instruments},
  volume={91},
  number={9},
  year={2020},
  publisher={AIP Publishing},
  doi = {10.1063/5.0017378}
}

@article{baselmans2012,
  title={Kinetic inductance detectors},
  author={Baselmans, Jochem},
  journal={Journal of Low Temperature Physics},
  volume={167},
  pages={292--304},
  year={2012},
  publisher={Springer},
  doi = {10.1007/s10909-011-0448-8}
}

@article{chiesa2023,
  title={Blueprint for a Molecular-Spin Quantum Processor},
  author={Chiesa, A and Roca, S and Chicco, S and de Ory, MC and G{\'o}mez-Le{\'o}n, A and Gomez, A and Zueco, D and Luis, F and Carretta, S},
  journal={Physical Review Applied},
  volume={19},
  number={6},
  pages={064060},
  year={2023},
  publisher={APS},
  doi = {10.1103/PhysRevApplied.19.064060}
}

@article{tabuchi2014,
  title={Hybridizing ferromagnetic magnons and microwave photons in the quantum limit},
  author={Tabuchi, Yutaka and Ishino, Seiichiro and Ishikawa, Toyofumi and Yamazaki, Rekishu and Usami, Koji and Nakamura, Yasunobu},
  journal={Physical review letters},
  volume={113},
  number={8},
  pages={083603},
  year={2014},
  publisher={APS},
  doi = {10.1103/PhysRevLett.113.083603}
}

@article{pirro2021,
  title={Advances in coherent magnonics},
  author={Pirro, Philipp and Vasyuchka, Vitaliy I and Serga, Alexander A and Hillebrands, Burkard},
  journal={Nature Reviews Materials},
  volume={6},
  number={12},
  pages={1114--1135},
  year={2021},
  publisher={Nature Publishing Group UK London},
  doi = {10.1038/s41578-021-00332-w}
}

@article{verjauw2021,
  title={Investigation of microwave loss induced by oxide regrowth in high-Q niobium resonators},
  author={Verjauw, J and Poto{\v{c}}nik, A and Mongillo, M and Acharya, R and Mohiyaddin, F and Simion, G and Pacco, A and Ivanov, Ts and Wan, D and Vanleenhove, A and others},
  journal={Physical Review Applied},
  volume={16},
  number={1},
  pages={014018},
  year={2021},
  publisher={APS},
  doi = {10.1103/PhysRevApplied.16.014018}
}

@article{gueron1996,
  title={Superconducting proximity effect probed on a mesoscopic length scale},
  author={Gu{\'e}ron, S and Pothier, H and Birge, Norman O and Esteve, D and Devoret, MH},
  journal={Physical review letters},
  volume={77},
  number={14},
  pages={3025},
  year={1996},
  publisher={APS},
  doi = {10.1103/PhysRevLett.77.3025}
}

@article{hu2020,
  title={Proximity-coupled Al/Au bilayer kinetic inductance detectors},
  author={Hu, Jie and Salatino, Maria and Traini, Alessandro and Chaumont, Christine and Boussaha, Faouzi and Goupil, Christophe and Piat, Michel},
  journal={Journal of Low Temperature Physics},
  volume={199},
  number={1-2},
  pages={355--361},
  year={2020},
  publisher={Springer},
  doi = {10.1007/s10909-019-02313-4}
}

@article{valenti2019,
  title={Interplay between kinetic inductance, nonlinearity, and quasiparticle dynamics in granular aluminum microwave kinetic inductance detectors},
  author={Valenti, Francesco and Henriques, Fabio and Catelani, Gianluigi and Maleeva, Nataliya and Gr{\"u}nhaupt, Lukas and von L{\"u}pke, Uwe and Skacel, Sebastian T and Winkel, Patrick and Bilmes, Alexander and Ustinov, Alexey V and others},
  journal={Physical review applied},
  volume={11},
  number={5},
  pages={054087},
  year={2019},
  publisher={APS},
  doi = {10.1103/PhysRevApplied.11.054087}
}

@article{chien2023,
  title={Large parametric amplification in kinetic inductance dominant resonators based on 3 nm-thick epitaxial superconductors},
  author={Chien, Wei-Chen and Chang, Yu-Han and Lu, Cheng Xin and Ting, Yen-Yu and Wu, Cen-Shawn and Lin, Sheng-Di and Kuo, Watson},
  journal={Materials for Quantum Technology},
  volume={3},
  number={2},
  pages={025005},
  year={2023},
  publisher={IOP Publishing},
  doi = {10.1088/2633-4356/acd744}
}

@article{samkharadze2016,
  title={High-kinetic-inductance superconducting nanowire resonators for circuit QED in a magnetic field},
  author={Samkharadze, Nodar and Bruno, A and Scarlino, Pasquale and Zheng, G and DiVincenzo, DP and DiCarlo, L and Vandersypen, LMK},
  journal={Physical Review Applied},
  volume={5},
  number={4},
  pages={044004},
  year={2016},
  publisher={APS},
  doi = {10.1103/PhysRevApplied.5.044004}
}

@article{cornia2011,
  title={Chemical strategies and characterization tools for the organization of single molecule magnets on surfaces},
  author={Cornia, Andrea and Mannini, Matteo and Sainctavit, Philippe and Sessoli, Roberta},
  journal={Chemical Society Reviews},
  volume={40},
  number={6},
  pages={3076--3091},
  year={2011},
  publisher={Royal Society of Chemistry},
  doi = {10.1039/C0CS00187B}
}

@article{gabarro2023,
  title={Magnetic molecules on surfaces: SMMs and beyond},
  author={Gabarr{\'o}-Riera, Guillem and Arom{\'\i}, Guillem and Sa{\~n}udo, E Carolina},
  journal={Coordination Chemistry Reviews},
  volume={475},
  pages={214858},
  year={2023},
  publisher={Elsevier},
  doi = {10.1016/j.ccr.2022.214858}
}

@article{berti2023,
  title={Scanning tunneling microscopy and spectroscopy characterization of Nb films for quantum applications},
  author={Berti, G and Torres-Castanedo, CG and Goronzy, DP and Bedzyk, MJ and Hersam, MC and Kopas, C and Marshall, J and Iavarone, M},
  journal={Applied Physics Letters},
  volume={122},
  number={19},
  year={2023},
  publisher={AIP Publishing},
  doi = {10.1063/5.0145090}
}

@article{wang2023universal,
  title={Universal quantum control of an atomic spin qubit on a surface},
  author={Wang, Yu and Haze, Masahiro and Bui, Hong T and Soe, We-hyo and Aubin, Herve and Ardavan, Arzhang and Heinrich, Andreas J and Phark, Soo-hyon},
  journal={npj Quantum Information},
  volume={9},
  number={1},
  pages={48},
  year={2023},
  publisher={Nature Publishing Group UK London},
  doi = {10.1038/s41534-023-00716-6}
}

@article{ghirri2011,
  title={Self-assembled monolayer of Cr7Ni molecular nanomagnets by sublimation},
  author={Ghirri, Alberto and Corradini, Valdis and Bellini, Valerio and Biagi, Roberto and Del Pennino, Umberto and De Renzi, Valentina and Cezar, Julio C and Muryn, Christopher A and Timco, Grigore A and Winpenny, Richard EP and others},
  journal={ACS nano},
  volume={5},
  number={9},
  pages={7090--7099},
  year={2011},
  publisher={ACS Publications},
  doi = {10.1021/nn201800e}
}

@MISC{sonnet,
   author       = {},
   title        = {Sonnet User’s Guide, Release 18}, 
   howpublished = {\url{https://www.sonnetsoftware.com/support}},
}

@article{probst2015,
  title={Efficient and robust analysis of complex scattering data under noise in microwave resonators},
  author={Probst, Sebastian and Song, FB and Bushev, Pavel A and Ustinov, Alexey V and Weides, Martin},
  journal={Review of Scientific Instruments},
  volume={86},
  number={2},
  year={2015},
  publisher={AIP Publishing},
  doi = {10.1063/1.4907935}
}

@article{day2003,
  title={A broadband superconducting detector suitable for use in large arrays},
  author={Day, Peter K and LeDuc, Henry G and Mazin, Benjamin A and Vayonakis, Anastasios and Zmuidzinas, Jonas},
  journal={Nature},
  volume={425},
  number={6960},
  pages={817--821},
  year={2003},
  publisher={Nature Publishing Group UK London},
  doi = {10.1038/nature02037}
}

@article{barends2009frequency,
  title={Frequency and quality factor of NbTiN/Au bilayer superconducting resonators},
  author={Barends, R and Daalman, WK-G and Endo, A and Zhu, S and Zijlstra, T and Klapwijk, TM},
  journal={AIP Conference Proceedings},
  volume={1185},
  number={1},
  pages={152--155},
  year={2009},
  organization={American Institute of Physics},
  doi = {10.1063/1.3292303}
}

@article{swenson2013,
  title={Operation of a titanium nitride superconducting microresonator detector in the nonlinear regime},
  author={Swenson, LJ and Day, PK and Eom, BH and Leduc, HG and Llombart, N and McKenney, CM and Noroozian, O and Zmuidzinas, J},
  journal={Journal of Applied Physics},
  volume={113},
  number={10},
  year={2013},
  publisher={AIP Publishing},
  doi = {10.1063/1.4794808}
}

@article{mattis1958,
  title={Theory of the anomalous skin effect in normal and superconducting metals},
  author={Mattis, Daniel C and Bardeen, John},
  journal={Physical Review},
  volume={111},
  number={2},
  pages={412},
  year={1958},
  publisher={APS},
  doi = {10.1103/PhysRev.111.412}
}

@article{zmuidzinas2012,
  title={Superconducting microresonators: Physics and applications},
  author={Zmuidzinas, Jonas},
  journal={Annu. Rev. Condens. Matter Phys.},
  volume={3},
  number={1},
  pages={169--214},
  year={2012},
  publisher={Annual Reviews},
  doi = {10.1146/annurev-conmatphys-020911-125022}
}

@article{vissers2015,
  title={Frequency-tunable superconducting resonators via nonlinear kinetic inductance},
  author={Vissers, Michael R and Hubmayr, Johannes and Sandberg, Martin and Chaudhuri, Saptarshi and Bockstiegel, Clint and Gao, Jiansong},
  journal={Applied Physics Letters},
  volume={107},
  number={6},
  year={2015},
  publisher={AIP Publishing},
  doi = {10.1063/1.4927444}
}

@article{andersson2021,
  title={Acoustic spectral hole-burning in a two-level system ensemble},
  author={Andersson, Gustav and Bilobran, Andr{\'e} Luiz Oliveira and Scigliuzzo, Marco and de Lima, Mauricio M and Cole, Jared H and Delsing, Per},
  journal={npj Quantum Information},
  volume={7},
  number={1},
  pages={15},
  year={2021},
  publisher={Nature Publishing Group UK London},
  doi = {10.1038/s41534-020-00348-0}
}

@article{burnett2017low,
  title={Low-loss superconducting nanowire circuits using a neon focused ion beam},
  author={Burnett, Jonathan and Sagar, James and Kennedy, Oscar W and Warburton, Paul A and Fenton, Jonathan C},
  journal={Physical Review Applied},
  volume={8},
  number={1},
  pages={014039},
  year={2017},
  publisher={APS},
  doi = {10.1103/PhysRevApplied.8.014039}
}

@article{frasca2023nbn,
  title={NbN films with high kinetic inductance for high-quality compact superconducting resonators},
  author={Frasca, Simone and Arabadzhiev, Ivo Nikolaev and de Puechredon, SY Bros and Oppliger, Fabian and Jouanny, Vincent and Musio, Roberto and Scigliuzzo, Marco and Minganti, Fabrizio and Scarlino, Pasquale and Charbon, Edoardo},
  journal={Physical Review Applied},
  volume={20},
  number={4},
  pages={044021},
  year={2023},
  publisher={APS},
  doi = {10.1103/PhysRevApplied.20.044021}
}

@article{bal2024systematic,
  title={Systematic improvements in transmon qubit coherence enabled by niobium surface encapsulation},
  author={Bal, Mustafa and Murthy, Akshay A and Zhu, Shaojiang and Crisa, Francesco and You, Xinyuan and Huang, Ziwen and Roy, Tanay and Lee, Jaeyel and Zanten, David van and Pilipenko, Roman and others},
  journal={npj Quantum Information},
  volume={10},
  number={1},
  pages={43},
  year={2024},
  publisher={Nature Publishing Group UK London},
  doi = {10.1038/s41534-024-00840-x}
}

@article{trasobares2023hybrid,
  title={Hybrid molecular graphene transistor as an operando and optoelectronic platform},
  author={Trasobares, Jorge and Mart{\'\i}n-Romano, Juan Carlos and Khaliq, Muhammad Waqas and Ruiz-G{\'o}mez, Sandra and Foerster, Michael and Ni{\~n}o, Miguel {\'A}ngel and Pedraz, Patricia and Dappe, Yannick J and de Ory, Marina Calero and Garc{\'\i}a-P{\'e}rez, Julia and others},
  journal={Nature Communications},
  volume={14},
  number={1},
  pages={1381},
  year={2023},
  publisher={Nature Publishing Group UK London},
  doi = {10.1038/s41467-023-36714-7}
}

@article{coronado2005isolated,
  title={Isolated Mn12 single-molecule magnets grafted on gold surfaces via electrostatic interactions},
  author={Coronado, Eugenio and Forment-Aliaga, Alicia and Romero, Francisco M and Corradini, Valdis and Biagi, Roberto and De Renzi, Valentina and Gambardella, Alessandro and Del Pennino, Umberto},
  journal={Inorganic chemistry},
  volume={44},
  number={22},
  pages={7693--7695},
  year={2005},
  publisher={ACS Publications}, 
  doi = {10.1021/ic0508021}
}

@article{alghadeer2022surface,
  title={Surface Passivation of Niobium Superconducting Quantum Circuits Using Self-Assembled Monolayers},
  author={Alghadeer, Mohammed and Banerjee, Archan and Hajr, Ahmed and Hussein, Hussein and Fariborzi, Hossein and Rao, Saleem Ghaffar},
  journal={ACS applied materials \& interfaces},
  volume={15},
  number={1},
  pages={2319--2328},
  year={2022},
  publisher={ACS Publications},
  doi = {10.1021/acsami.2c15667}
}
\end{multicols}
\end{document}